# Energy Based Equality of Distributions Testing for Compositional Data

**Volkan Sevinç**[1] **Michail Tsagris**[2]

[1]Department of Statistics, Faculty of Science, Muğla Sıtkı Koçman University, Kötekli Kampüsü, 48000, Muğla, Türkiye, vsevinc@mu.edu.tr

[2]Department of Economics, School of Social Sciences, University of Crete, 74100, Rethymno, Crete, Greece, mtsagris@uoc.gr

## ABSTRACT

In the literature, there are different kinds of tests for testing the equality of two or more multivariate distributions. However, these tests cannot be used for comparing the distributions of compositional data sets due to their special structures and constrained sample spaces. In the current literature, for comparison of such distributions, there is only one proposed test, which is based on random projections. Therefore, this article proposes a novel non-parametric test called $\alpha$–Energy Based Test ($\alpha$–EBT) to compare the multivariate distributions of two or more compositional data sets. To compare the testing performances of the random projections based test and the newly proposed $\alpha$–EBT, simulation studies and an application, which uses real data from agricultural economics are also provided within the frame of the study. The application and the simulation study results show that $\alpha$–EBT demonstrates higher power levels and can successfully be used for comparing compositional data distributions.

## 1. Introduction

A compositional data vector can be described as a special vector type of multivariate observations, in which the elements are non-negative and their sum is equal to a constant, typically chosen to be 1. The related sample space of this vector is the standard simplex, defined as

$$\mathbb{S}^{D-1} = \{(x_1, \dots, x_D)^\top | x_i \geq 0, \textstyle\sum_{i=1}^{D} x_i = 1\}, \tag{1}$$

where $D$ denotes the number of variables (also known as components).

Compositional data are used in many different scientific fields. There are many studies that have been published regarding the analyses and real-life applications of compositional data[1].

There are three main approaches used for analyzing compositional data. The first one, is the stay-in-the-simplex approach that employs distributions and models defined on the simplex. Dirichlet distribution and related models (Ng et al., 2011) is one such example. Another example is Iyengar and Dey (2002) who examined the generalized Liouville distribution family, which allows negative or mixed correlations and extends beyond Dirichlet distributions to include non-positive correlation structures. The second one, called raw data analysis (RDA), neglects the compositional constraint and applies the standard multivariate data analysis techniques to the raw data. For discussion and examples of RDA, Baxter (1995, 2001); Baxter et al. (2005); Baxter and Freestone (2006) may be referred. The third, and most popular approach, is the log-ratio analysis (LRA) introduced by Aitchison (1982, 1983, 1986). Aitchison advocated that the idea behind the LRA is that since compositional data carry only relative information about the components, log-ratio transformations are quite suitable for analyzing this information.

However, there is a problem with the LRA approach, which is the possible presence of zero values making the logarithmic transformation impossible. Although various zero replacement strategies (Aitchison, 1986) have been suggested in the literature to address this problem, such a procedure can introduce measurement errors into the data and thus, may cause bias in the statistical analysis.

Therefore, when the distributions of two or more compositional data sets are to be compared, the standard comparison tests are difficult or sometimes impossible to apply due to the problems explained above. Other approaches involving other kinds of transformations need to be employed for comparison of compositional data distributions. Aitchison (1986) proposed a collection of tests for testing the normality of compositional data sets; however, these tests are solely devised for the multivariate normal distribution. Cuesta-Albertos et al. (2009) were the first to address this problem in the two-sample case. They proposed a non-parametric, random projections based test (RPBT) for testing the equality of distributions of two compositional data sets. However, even in their small-scale simulation studies, it was evident that the test was not right-sized.

[1] For an indicative list of fields where compositional data are frequently met, see Tsagris and Stewart (2020).

**Research gap and motivation of the study**

Due to the reasons explained above, the standard goodness-of-fit tests cannot be used for comparison of compositional data distributions. Moreover, the only comparison test in the current literature proposed for compositional data has some shortages. Therefore, we fill this gap by developing a new test for compositional data. The test we propose is achieved through the utilization of the $\alpha$–transformation (Tsagris et al., 2011), a power transformation that offers an extra degree of flexibility in compositional data analysis. We applied the $\alpha$–transformation to the non-parametric test called Energy Based Test (EBT), which was proposed by Székely and Rizzo (2004). Therefore, we call the resulting test the $\alpha$–Energy Based Test ($\alpha$–EBT), which overcomes the problems arising from the use of compositional data. The ordinary EBT relies upon distances of the observations. Hence, it could be argued that it could be directly applied to the compositional data without any transformations regardless of the presence of zero values. However, the necessity of the transformation is a reality, which will be made evident in the simulation studies.

The next section reviews some transformations for compositional data, while Section 3 reviews the RPBT, the EBT, and introduces the $\alpha$–EBT. Section 4 contains extensive simulation studies comparing the two tests in terms of Type I and Type II errors. Section 5 contains an example with real data coming from the field of agricultural economics. Finally, the last section concludes the paper.

## 2. Transformations for Compositional Data

The Dirichlet distribution is a naturally appropriate distribution for compositional data as the support of this distribution is the simplex space. On the other hand, it is known that this distribution is not statistically rich and flexible enough to capture the different kinds of variabilities including the curvature of compositional data. Therefore, different types of transformations have been suggested to map the data outside the simplex. Among them, Aitchison (1982) suggested the log-ratio transformation approach, and later the isometric log-ratio (ILR) transformation[2] methodology was proposed and examined in detail by Egozcue et al. (2003). Moreover, the $\alpha$–transformation, which is a Box−Cox type transformation and includes the ILR transformation as a special case, was suggested by Tsagris et al. (2011). The $\alpha$–transformation has been successfully employed in regression analysis (Tsagris 2015) and classification studies (Tsagris 2014; Tsagris et al. 2016).

---

[2] This transformation was first mentioned by Aitchison (1986, pg. 90).

### 2.1. Log-ratio transformations

The first log-ratio transformation, which was suggested by Aitchison (1983, 1986), is termed the centered log-ratio (CLR) transformation and given by

$$\mathbf{w}_0(\mathbf{x}) = \left(\log\left(\frac{x_1}{\prod_{j=1}^{D} x_j^{1/D}}\right), \dots, \log\left(\frac{x_D}{\prod_{j=1}^{D} x_j^{1/D}}\right)\right)^{\top}, \quad (2)$$

where the component $\left(\prod_{j=1}^{D} x_j\right)^{1/D}$ is the geometric mean of the $D$ components of the compositional data. The sample space of (2) is given by

$$\mathbb{Q}_0^{D-1} = \left\{\left(w_{0,1}, \dots, w_{0,\mathrm{D}}\right)^{\top} \middle| \sum_{i=1}^{D} w_{0,i} = 0\right\}, \quad (3)$$

which is a subset of $\mathbb{R}^{D-1}$. The zero-sum constraint in (3) is obviously a disadvantage of this transformation because it may cause singularity issues. To remove the redundant dimension caused by this constraint, one can left multiply (2) by an orthonormal matrix yielding the ILR transformation defined as follows.

$$\mathbf{z}_0(\mathbf{x}) = \mathbf{H}\mathbf{w}_0(\mathbf{x}), \quad (4)$$

where $\mathbf{z}_0(\mathbf{x})$ is a $D-1$ dimensional vector and $\mathbf{H}$ is the $(D-1) \times D$ Helmert sub-matrix (Lancaster, 1965), i.e. the Helmert matrix after deleting the first row. This sub-matrix is a standard orthogonal matrix in shape analysis and used to overcome the singularity problems[3]. The sample space of (4) is $\mathbb{R}^{D-1}$ because left multiplication by the Helmert sub-matrix maps the CLR transformed data onto $\mathbb{R}^{D-1}$, so that the zero-sum constraint is removed.

### 2.2. The $\alpha$–transformation

Tsagris et al. (2011) developed the $\alpha$–transformation as a more general transformation than that in (4). Let

$$\mathbf{u}_\alpha(\mathbf{x}) = \left(\frac{x_1^{\alpha}}{\sum_{j=1}^{D} x_j^{\alpha}}, \dots, \frac{x_D^{\alpha}}{\sum_{j=1}^{D} x_j^{\alpha}}\right)^{\top} \quad (5)$$

denote the power transformation for compositional data as defined by Aitchison (1986), where $\alpha$ can take any value between $-1$ and $1$ (when zero values exist in the data, $\alpha$ can take only strictly positive values). In a manner analogous to (2) $-$ (4), first define

[3] For further information, see Dryden and Mardia (1998); Le and Small (1999).

$$\mathbf{w}_\alpha(\mathbf{x}) = \frac{D\mathbf{u}_\alpha(\mathbf{x})-1}{\alpha}. \qquad (6)$$

The sample space of (6) is

$$\mathbb{Q}_\alpha^{D-1} = \left\{ \left(w_{\alpha,1}, \ldots, w_{\alpha,D}\right)^\top \middle| \frac{-1}{\alpha} \leqslant w_{\alpha,i} \leqslant \frac{D-1}{\alpha}, \sum_{i=1}^D w_{\alpha,i} = 0 \right\}. \qquad (7)$$

As $\alpha \to 0$ (6) converges to (2), and (7) becomes equal to (3). Finally, the $\alpha$–transformation is defined as

$$\mathbf{z}_\alpha(\mathbf{x}) = \mathbf{H}\mathbf{w}_\alpha(\mathbf{x}). \qquad (8)$$

Similar to the ILR transformation, the purpose of $\mathbf{H}$ is to remove the redundant dimension arising from the compositional constraint. In particular, the vector $(D\mathbf{u}_\alpha(\mathbf{x}) - \mathbf{1}_D)/\alpha$ has components which sum to zero and thereby exist in a subspace of $\mathbb{R}^D$. Left multiplication by $\mathbf{H}$ is an isometric one-to-one mapping from this subspace into $\mathbb{R}^{D-1}$. Further, note that any orthonormal matrix with similar properties to $\mathbf{H}$ may be used. The $\alpha$–transformation is a one-to-one transformation, which maps the data from the simplex onto a subset of $\mathbb{R}^{D-1}$ and vice versa for $\alpha \neq 0$, and its corresponding sample space is as follows.

$$\mathbb{A}_\alpha^{D-1} = \left\{ \mathbf{H}\mathbf{w}_\alpha(\mathbf{x}) \middle| -\frac{1}{\alpha} \leqslant w_{\alpha,i} \leqslant \frac{D-1}{\alpha}, \sum_{i=1}^D w_{\alpha,i} = 0 \right\}. \qquad (9)$$

Note that vectors in $\mathbb{A}_\alpha^{D-1}$ are not subject to the sum to zero constraint and that $\lim_{\alpha\to 0} \mathbb{A}_\alpha^{D-1} = \mathbb{R}^{D-1}$. Further, for (4) and (5), when $\alpha = 1$, the simplex is linearly expanded as the values of the components are simply multiplied by a scalar, and then centered. When $\alpha = -1$, the inverse of the values of the components are multiplied by a scalar and then centered. The $\alpha$–transformation, (8) in the limiting case $\alpha \to 0$ converges to the ILR transformation, which corresponds to the ILR transformation (4) and hence to LRA, whereas $\alpha = 1$ corresponds to RDA.

The $\alpha$–transformation (8) leads to a natural simplicial distance measure $\Delta_\alpha(\mathbf{x}, \mathbf{y})$, which we call the $\alpha$–metric, between the observations $\mathbf{x}, \mathbf{y} \in \mathbb{S}^d$, defined in terms of the Euclidean distance $\|\cdot\|$ between transformed observations, i.e.,

$$\Delta_\alpha(\mathbf{x}, \mathbf{y}) = \|\mathbf{w}_\alpha(\mathbf{x}) - \mathbf{w}_\alpha(\mathbf{y})\| = \frac{D}{|\alpha|} \left[ \sum_{i=1}^D \left( \frac{x_i^\alpha}{\sum_{j=1}^D x_j^\alpha} - \frac{y_i^\alpha}{\sum_{j=1}^D y_j^\alpha} \right)^2 \right]^{1/2}. \qquad (10)$$

The special case

$$\Delta_0(\mathbf{x}, \mathbf{y}) := \lim_{\alpha\to 0} \Delta_\alpha(\mathbf{x}, \mathbf{y}) = \left[ \sum_{i=1}^D \left( \log \frac{x_i}{g(\mathbf{x})} - \log \frac{y_i}{g(\mathbf{y})} \right)^2 \right]^{1/2} \qquad (11)$$

is Aitchison's distance measure (Aitchison et al., 2000), whereas

$$\Delta_1(\mathbf{x}, \mathbf{y}) = D\left[\sum_{i=1}^{D} (x_i - y_i)^2\right]^{1/2} \tag{12}$$

is simply the Euclidean distance applied to the raw (untransformed) data, multiplied by $D$.

The distance measure seen in (10) presents flexibility in data analysis, that is, the choice of $\alpha$ enables either LRA or RDA, or a compromise between them. In addition, the value of $\alpha$ can be chosen to optimize some measure of practical performance (in this paper, the power of the EBT test). Crucially, for $\alpha > 0$, the transformation and distance are well defined even when some components have zero values, in contrast to (2), (4), and (11).

## 3. Tests of Equality of Simplicial Distributions

### 3.1. The RPBT

For comparison of compositional data sets, Cuesta-Albertos et al. (2009) proposed the RPBT, which is a non-parametric test based on random projections. Since their test was devised for directional data, the compositional vectors first must be transformed to unit vectors by computing the square root of each element. In a two-sample problem involving compositional data, the RPBT concludes that both samples belong to the same distribution, if their respective one-dimensional projections along a random direction are identically distributed. The latter can be tested via the non-parametric Kolmogorov−Smirnov (KS) test.

Since many random projections are to be made, Cuesta-Albertos et al. (2009) proposed to use the Bonferroni procedure to combine the tests performed along the different projections, instead of using the test based on the minimum $p-$value. Again, the reason is that the null hypothesis assuming homogeneity is composite, and in general, the null distribution of the minimum $p-$value depends on the common distribution of both samples. Still, the resulting $p-$values are correlated, and for this reason, we will use the Benjamini and Heller (2008). Suppose $B$ projections were performed and hence $B$ $p-$values, i.e. $p_i, i = 1, \dots, B$ were produced from the relevant KS tests. At first, sort the $p-$values, $p_{(1)} < \cdots < p_{(B)}$ and compute the combined $p-$value as $p-$value $= \min_i \left\{\frac{B}{i} p_{(i)}\right\}$.

### 3.2. The EBT

Székely and Rizzo (2004) proposed the nonparametric test EBT for testing the equality of two or more multivariate distributions. The EBT is based on the Euclidean distance between sample elements, and implemented by conditioning on the pooled sample to obtain a distribution-free approximate permutation test. Székely and Rizzo (2004) describe the fundamentals of the EBT as follows.

To introduce the $e$-distance between finite sets, let $\mathcal{A} = \{a_1, \dots, a_{n_1}\}$ and $B = \{b_1, \dots, b_{n_2}\}$ be disjoint nonempty subsets of $\mathbb{R}^d$ where $n_1$ and $n_2$ denote the size of the sets. The $e$-distance $e(\mathcal{A}, B)$ between $\mathcal{A}$ and $B$ is

$$e(\mathcal{A}, B) = \frac{n_1 n_2}{n_1+n_2}\left(\frac{2}{n_1 n_2}\sum_{i=1}^{n_1}\sum_{j=1}^{n_2}\left\|a_i - b_j\right\| - \frac{1}{n_1^2}\sum_{i=1}^{n_1}\sum_{j=1}^{n_2}\left\|a_i - a_j\right\| - \frac{1}{n_2^2}\sum_{i=1}^{n_1}\sum_{j=1}^{n_2}\left\|b_i - b_j\right\|\right). \quad (13)$$

Let $\boldsymbol{X}_1, \dots, \boldsymbol{X}_{n_1}$ and $\boldsymbol{Y}_1, \dots, \boldsymbol{Y}_{n_2}$ be independent random samples of random vectors in $\mathbb{R}^D$, $D \geq 1$. If $\mathcal{A}$ denotes the first sample, and $\mathcal{B}$ denotes the second sample, with corresponding sample sizes $n_1$ and $n_2$, the two-sample test statistic $EBT_{n_1,n_2}$ is given as follows

$$EBT_{n_1,n_2} = \frac{n_1 n_2}{n_1+n_2}\left(\frac{2}{n_1 n_2}\sum_{i=1}^{n_1}\sum_{m=1}^{n_2}\|\boldsymbol{X}_i - \boldsymbol{Y}_m\| - \frac{1}{n_1^2}\sum_{i=1}^{n_1}\sum_{j=1}^{n_1}\left\|\boldsymbol{X}_i - \boldsymbol{X}_j\right\| - \frac{1}{n_2^2}\sum_{l=1}^{n_2}\sum_{m=1}^{n_2}\|\boldsymbol{Y}_l - \boldsymbol{Y}_m\|\right). \quad (14)$$

The distribution of the test statistic (14) under the null hypothesis is unknown and hence, the $p-$value of the test is approximated via permutations.

The computational complexity of the energy distance and hence of the EBT is $O(n^2 D)$. The implementation of the EBT in the *R* package *energy* (Rizzo and Szekely, 2024) is prohibitive with large sample sized data. We have implemented a memory efficient version of the EBT test in the *R* package *Compositional* (Tsagris et al., 2025). Eq. (14) may be simplified (excluding the terms that do not alter during permutations) as

$$EBT_{n_1,n_2} = \frac{2}{n_1+n_2}\delta_{xy} - \frac{n_1}{n_1+n_2}\delta_x - \frac{n_2}{n_1+n_2}\delta_y,$$

*where* $\delta_{xy} = \sum_{i=1}^{n_1}\sum_{m=1}^{n_2}\|\boldsymbol{X}_i - \boldsymbol{Y}_m\|$, $\delta_x = \sum_{i=1}^{n_1}\sum_{j=1}^{n_1}\left\|\boldsymbol{X}_i - \boldsymbol{X}_j\right\|$, and $\delta_y = \sum_{l=1}^{n_2}\sum_{m=1}^{n_2}\|\boldsymbol{Y}_l - \boldsymbol{Y}_m\|$.

The terms $\delta_x$ and $\delta_y$ are computed using the command *Dist(x, result = "sum")* and *Dist(y, result = "sum")*, respectively from the *Rfast* package (Papadakis et al., 2025). The term $\delta_{xy}$ is computed using the command dista(x, y, result = "sum") from the same package. These computations require no extra memory at all, rendering the computation of the EBT highly memory efficient. The sum of the three terms $\delta = \delta_{xy} + \delta_x + \delta_y$ is constant. Hence, during the permutations one needs to compute the two out of the three terms, and the choice between $\delta_x$ and $\delta_y$ is to compute the one term that corresponds to the sample with the smallest sample size.

### 3.3. The proposed $\alpha$–EBT

Our proposal relies upon substituting the sample vectors in (14) with the $\alpha$–transformed compositional data, $\mathbf{w}_\alpha(\mathbf{x})$ and $\mathbf{w}_\alpha(\mathbf{y})$, substituting the Euclidean distances with the $\alpha$–metrics between the compositional vectors of the data sets. Thus, the proposed $\alpha$–$EBT_{n_1,n_2}$ statistic is given as follows.

$$
\begin{aligned}
\alpha - EBT_{n_1,n_2} &= \frac{n_1 n_2}{n_1+n_2}\Bigg(\frac{2}{n_1 n_2}\sum_{l=1}^{n_1}\sum_{m=1}^{n_2}\Delta_\alpha(\mathbf{x}_l,\mathbf{y}_m) - \frac{1}{n_1^2}\sum_{l=1}^{n_1}\sum_{m=1}^{n_1}\Delta_\alpha(\mathbf{x}_l,\mathbf{x}_m) \\
&\quad - \frac{1}{n_2^2}\sum_{l=1}^{n_2}\sum_{m=1}^{n_2}\Delta_\alpha(\mathbf{y}_l,\mathbf{y}_m)\Bigg), \\
&= \frac{n_1 n_2}{n_1+n_2}\Bigg(\frac{2}{n_1 n_2}\sum_{l=1}^{n_1}\sum_{m=1}^{n_2}\|\mathbf{w}_\alpha(\mathbf{x}_l) - \mathbf{w}_\alpha(\mathbf{x}_m)\| - \frac{1}{n_1^2}\sum_{l=1}^{n_1}\sum_{m=1}^{n_1}\|\mathbf{w}_\alpha(\mathbf{x}_l) - \mathbf{w}_\alpha(\mathbf{x}_m)\| \\
&\quad - \frac{1}{n_2^2}\sum_{l=1}^{n_2}\sum_{m=1}^{n_2}\|\mathbf{w}_\alpha(\mathbf{y}_l) - \mathbf{w}_\alpha(\mathbf{y}_m)\|\Bigg), \\
&= \frac{D}{|\alpha|}\frac{n_1 n_2}{n_1+n_2}\Bigg\{\frac{2}{n_1 n_2}\sum_{l=1}^{n_1}\sum_{m=1}^{n_2}\left[\sum_{i=1}^{D}\left(\frac{x_{il}^{\alpha}}{\sum_{j=1}^{D}x_{lj}^{\alpha}} - \frac{y_{im}^{\alpha}}{\sum_{j=1}^{D}y_{jm}^{\alpha}}\right)^2\right]^{1/2} \\
&\quad - \frac{1}{n_1^2}\sum_{l=1}^{n_1}\sum_{m=1}^{n_1}\left[\sum_{i=1}^{D}\left(\frac{x_{il}^{\alpha}}{\sum_{j=1}^{D}x_{jl}^{\alpha}} - \frac{x_{im}^{\alpha}}{\sum_{j=1}^{D}x_{jm}^{\alpha}}\right)^2\right]^{1/2} \\
&\quad - \frac{1}{n_2^2}\sum_{l=1}^{n_2}\sum_{m=1}^{n_2}\left[\sum_{i=1}^{D}\left(\frac{y_{il}^{\alpha}}{\sum_{j=1}^{D}y_{jl}^{\alpha}} - \frac{y_{im}^{\alpha}}{\sum_{j=1}^{D}y_{jm}^{\alpha}}\right)^2\right]^{1/2}\Bigg\}. \qquad (15)
\end{aligned}
$$

The permutation-based $p-$value is again used to determine the rejection of the null hypothesis. However, this time, an extra degree of complexity encompasses the test, which is the choice of the value of $\alpha$. It was earlier mentioned that the possible values range within $[-1, 1]$, and when there are zero values in the data, the range reduces to strictly positive values of $\alpha$. Yet, the continuum of possible values is still wide.

Zero-inflated data are common in practice and may distort distance computations, and potentially inflating Type I error probability or reducing power. However, zero values in the observed compositional data, after the $\alpha$–transformation become $\frac{-1}{\alpha}$ and are then multiplied by the Helmert sub-matrix. Thus, the $\alpha$– transformed data contain no zeros. Further, the data are standardized so that all components are in the same scale removing the scale effect in the Euclidean distance.

Tsagris et al. (2017) considered the case of $\alpha = 1$, but they were interested in the simplicial means. However, we are interested in the simplicial distributions. To this end, we decided to use a range of values of $\alpha$ and see what evidence is suggested by each value.

## 4. Simulation Studies

The simulation studies were conducted to assess the performances of both tests in terms of Type I and Type II errors to see the effect of the $\alpha$ on the performance of the $\alpha$–EBT. The data were generated from the Dirichlet distribution, and the multivariate normal on the simplex, using the inverse of the additive log-ratio (ALR) transformation. Upon generation of a vector $\boldsymbol{u}$, it is mapped onto the simplex by using

$$\mathbf{x} = \left( \frac{1}{1+\sum_{j=2}^{D} e^{u_j}}, \frac{e^{u_2}}{1+\sum_{j=2}^{D} e^{u_j}} \dots, \frac{e^{u_D}}{1+\sum_{j=2}^{D} e^{u_j}} \right)^{\mathsf{T}}. \tag{16}$$

The sample sizes considered were $\mathrm{n} = (50, 100, 200, 300, 500, 1000)$ and the dimensionality of the simplex was set equal to $D = (3, 5, 10, 20, 30)$. Five distinct values of $\alpha = (0.1, 0.25, 0.5, 0.75, 1)$ were tested for the $\alpha$–EBT, but the obtained results rely only on two values of $\alpha$, which are 0.1 and 1, for illustration purposes. 100 projections were used for the RPBT and 999 permutations for the $\alpha$–EBT.

For all cases, 1,000 Monte Carlo iterations were performed to estimate the Type I error probability. The simulations were run on a Dell laptop with Intel Core i5-10351G1 CPU at 1.00GHz, 8GB RAM and Windows 10 installed.

The space of the $\alpha$–transformed data depends upon the value of $\alpha$, and as stated earlier, as it tends to zero, the space coincides with $\mathbb{R}^{D-1}$. This implies that the scale of the transformed data could influence the testing procedure. For this reason the experiments were performed twice. Once with the (variable-wise) standardized $\alpha$–transformed values before applying the $\alpha$–EBT, and once without standardization. However, it can be concluded that the results did not change even with the standardization process.

### 4.1. Type I error probability

Table 1 presents the estimated Type I error probabilities produced after 1,000 iterations, where the value of $\alpha$ in the $\alpha$–EBT was equal to 1. Evidently, only the $\alpha$–EBT attains the nominal correct size (5%) regardless of the size and dimensionality of the simplex. Note also, the value of $\alpha$ did not affect the Type I error probability. The RPBT on the contrary, is close to the nominal size only when $D = 20$ or $D = 30$ and the sample size is at least 100. Hence, the powers of both tests were assessed only for the case of $D = 30$.

**Table 1.** Estimated Type I error probabilities of the PBT and the $\alpha$–EBT (with $\alpha = 1$) when the data were generated from the Dirichlet and the simplicial multivariate normal.

| | **Dirichlet Distribution** | | | | | | | | | |
|---|---|---|---|---|---|---|---|---|---|---|
| | **D=3** | | **D=5** | | **D=10** | | **D=20** | | **D=30** | |
| | **RPBT** | **$\alpha$–EBT** | **RPBT** | **$\alpha$–EBT** | **RPBT** | **$\alpha$–EBT** | **RPBT** | **$\alpha$–EBT** | **RPBT** | **$\alpha$–EBT** |
| **n=50** | 0.021 | 0.051 | 0.016 | 0.035 | 0.025 | 0.060 | 0.027 | 0.050 | **0.020** | **0.054** |
| **n=100** | 0.024 | 0.056 | 0.027 | 0.046 | 0.044 | 0.061 | 0.034 | 0.052 | **0.036** | **0.049** |
| **n=200** | 0.023 | 0.048 | 0.036 | 0.052 | 0.031 | 0.048 | 0.051 | 0.067 | **0.042** | **0.059** |
| **n=300** | 0.015 | 0.044 | 0.032 | 0.053 | 0.037 | 0.058 | 0.038 | 0.048 | **0.039** | **0.046** |
| **n=500** | 0.021 | 0.054 | 0.029 | 0.048 | 0.032 | 0.047 | 0.046 | 0.055 | **0.039** | **0.039** |
| **n=1000** | 0.026 | 0.059 | 0.024 | 0.045 | 0.026 | 0.049 | 0.051 | 0.054 | **0.039** | **0.056** |
| | **Simplicial Multivariate Normal Distribution** | | | | | | | | | |
| | **D=3** | | **D=5** | | **D=10** | | **D=20** | | **D=30** | |
| | **RPBT** | **$\alpha$–EBT** | **RPBT** | **$\alpha$–EBT** | **RPBT** | **$\alpha$–EBT** | **RPBT** | **$\alpha$–EBT** | **RPBT** | **$\alpha$–EBT** |
| **n=50** | 0.021 | 0.048 | 0.027 | 0.053 | 0.032 | 0.051 | 0.029 | 0.052 | **0.027** | **0.050** |
| **n=100** | 0.024 | 0.050 | 0.039 | 0.060 | 0.030 | 0.052 | 0.031 | 0.050 | **0.038** | **0.044** |
| **n=200** | 0.027 | 0.050 | 0.029 | 0.049 | 0.022 | 0.038 | 0.041 | 0.051 | **0.034** | **0.039** |
| **n=300** | 0.030 | 0.049 | 0.025 | 0.049 | 0.037 | 0.058 | 0.031 | 0.052 | **0.043** | **0.047** |
| **n=500** | 0.026 | 0.056 | 0.035 | 0.055 | 0.033 | 0.042 | 0.031 | 0.044 | **0.036** | **0.049** |
| **n=1000** | 0.017 | 0.034 | 0.035 | 0.057 | 0.023 | 0.058 | 0.033 | 0.045 | **0.042** | **0.045** |

### 4.2. Power

Regarding the powers of the RPBT and the $\alpha$–EBT tests, the Dirichlet and the simplicial multivariate normal distributions were used to generate the data for the case of $D = 30$. In total, five scenarios were tested, one scenario using the Dirichlet distribution and four scenarios using the simplicial multivariate normal, as listed below:

- **Scenario 1**: The values of the first sample were generated from a $Dir(f \times \mathbf{3}_D)$, whereas the values of the second sample were generated from a $Dir(\mathbf{3}_D)$, where $\mathbf{3}_D = (3, \dots, 3)^T$ is a $D$-dimensional vector of 3s. The range of values of $f = (1, 1.1, \dots, 2)$ controls the divergence from the null hypothesis assuming equal distributions. When $f = 1$, the two distributions are the same, but as $f$ increases, so does the discrepancy between the two distributions.

- **Scenario 2**: The values of the first sample were generated from a $MN_{D-1}(f \times \boldsymbol{\mu}, \boldsymbol{\Sigma})$, whereas the values of the second sample were generated from $MN_{D-1}(\boldsymbol{\mu}, \boldsymbol{\Sigma})$, where in both cases $MN_{D-1}$ denotes the multivariate normal in $R^{D-1}$ with $\boldsymbol{\mu}$ generated from the standard normal distribution. The covariance matrix $\boldsymbol{\Sigma}$ was generated in the same manner as in Fayomi et al. (2024). At first, an orthonormal basis $\boldsymbol{B}$ was created using QR decomposition, then eigenvalues $\lambda_1, \ldots, \lambda_{D-1}$ were generated from an exponential distribution with mean equal to 0.4. Then, $\boldsymbol{\Sigma} = \boldsymbol{B}\, diag(\lambda_1, \ldots, \lambda_{D-1})\, \boldsymbol{B}^T$ and the values of $f$ are the same as before.
- **Scenario 3**: The values of the first sample were generated from a $MN_{D-1}(\boldsymbol{\mu}, \Sigma^*)$, where $\boldsymbol{\Sigma}^* = \boldsymbol{B}\, diag(f \times \lambda_1, \ldots, f \times \lambda_{D-1})\, \boldsymbol{B}^T$, and the values of the second sample were generated from $MN_{D-1}(\boldsymbol{\mu}, \boldsymbol{\Sigma})$.
- **Scenario 4**: The values of the first sample were generated from a $MN_{D-1}(f \times \boldsymbol{\mu}, \Sigma^*)$, where $\boldsymbol{\Sigma}^*$ was as defined in Scenario 3, and the values of the second sample were generated from $MN_{D-1}(\boldsymbol{\mu}, \boldsymbol{\Sigma})$.
- **Scenario 5**: The values of the first sample were generated from a $MN_{D-1}(\boldsymbol{\mu}, \boldsymbol{\Sigma}^*)$, where $\boldsymbol{\Sigma}^*$ was as defined in Scenario 3, and the values of the second sample were generated from $MN_{D-1}(f \times \boldsymbol{\mu}, \boldsymbol{\Sigma})$.

To effectively quantify the discrepancy between the true distributions of each sample, we computed the Kullback−Leibler (KL) divergence, whose formulas for the Dirichlet and the multivariate normal are given below.

$$KL[D_1(\boldsymbol{a})\,||D_2(\boldsymbol{b})] = \sum_{i=1}^{D}(a_i - b_i)\,[\psi(a_i) - \psi(a_0)] + \sum_{i=1}^{D} log\frac{\Gamma(b_i)}{\Gamma(a_i)} + log\frac{\Gamma(a_0)}{\Gamma(b_0)}, \tag{17}$$

where $a_0 = \sum_{i=1}^{D} a_i$, $b_0 = \sum_{i=1}^{D} b_i$ and $\psi(.)$ is the digamma function.

$$KL(MN_1\,||MN_2) = \frac{1}{2}\left[tr(\Sigma_2^{-1}\Sigma_1) + (\boldsymbol{\mu}_2 - \boldsymbol{\mu}_1)^T\Sigma_2^{-1}(\boldsymbol{\mu}_2 - \boldsymbol{\mu}_1) - log\frac{|\Sigma_1|}{|\Sigma_2|} - (D-1)\right]. \tag{18}$$

The estimated power of the RPBT and the $\alpha$–EBT as a function of the KL divergence under Scenario 1, 2, 3, 4, and 5 are presented in Figure 1, 2, 3, 4, and 5, respectively. The simulation studies (not shown here) indicated that the value of $\alpha$ had a significant effect on the estimated power level of the $\alpha$–EBT. Specifically, as the value of $\alpha$ decreased, the power increased. The figures contain the estimated powers of the $\alpha$–EBT test when $\alpha = 0.1$ and when $\alpha = 1$.

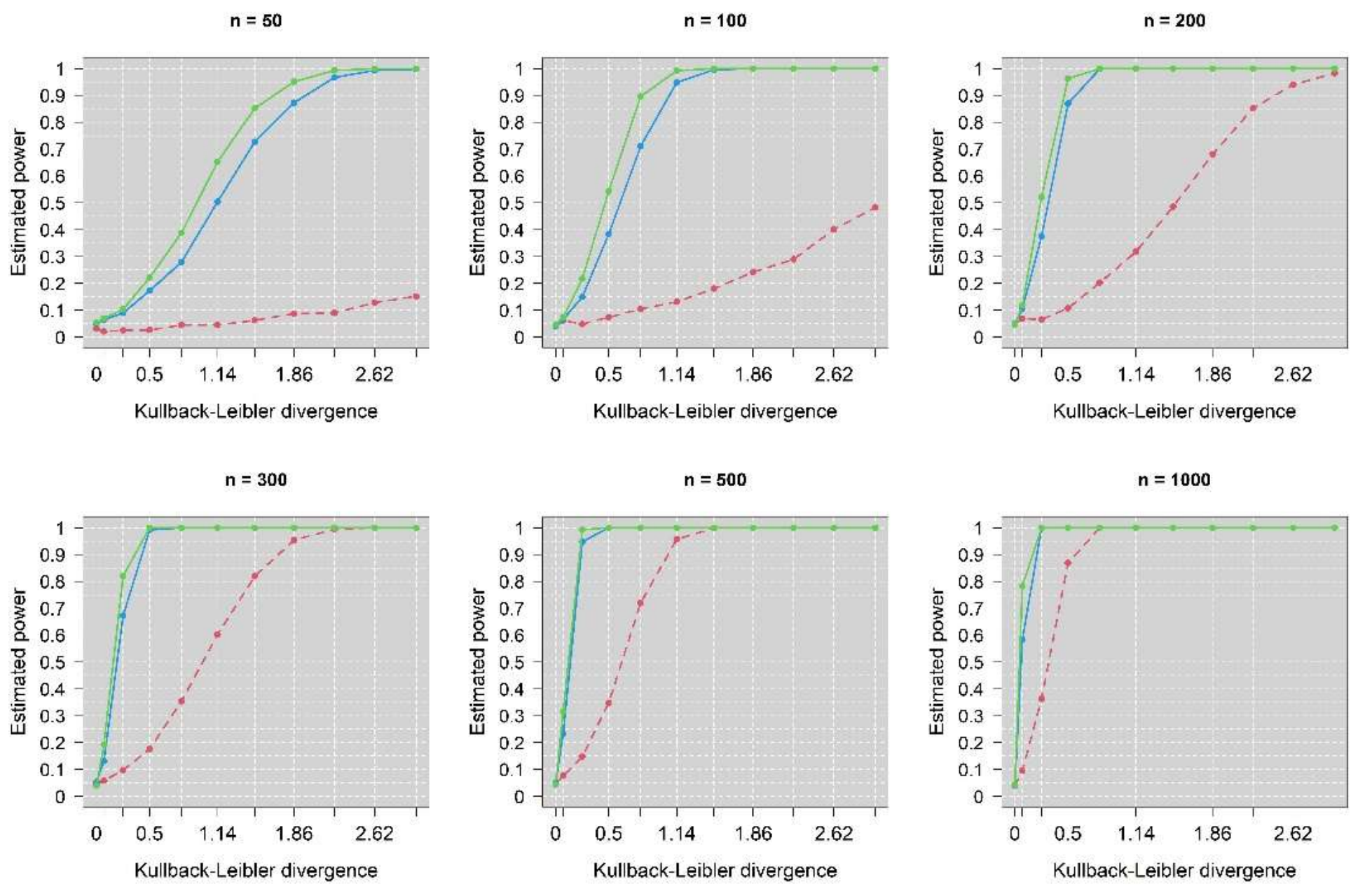

**Figure 1.** Estimated power of the RPBT (red line), and $\alpha$–EBT with $\alpha = 1$ (blue line), and $\alpha$–EBT with $\alpha = 0.1$ (green line) as a function of the KL divergence under Scenario 1.

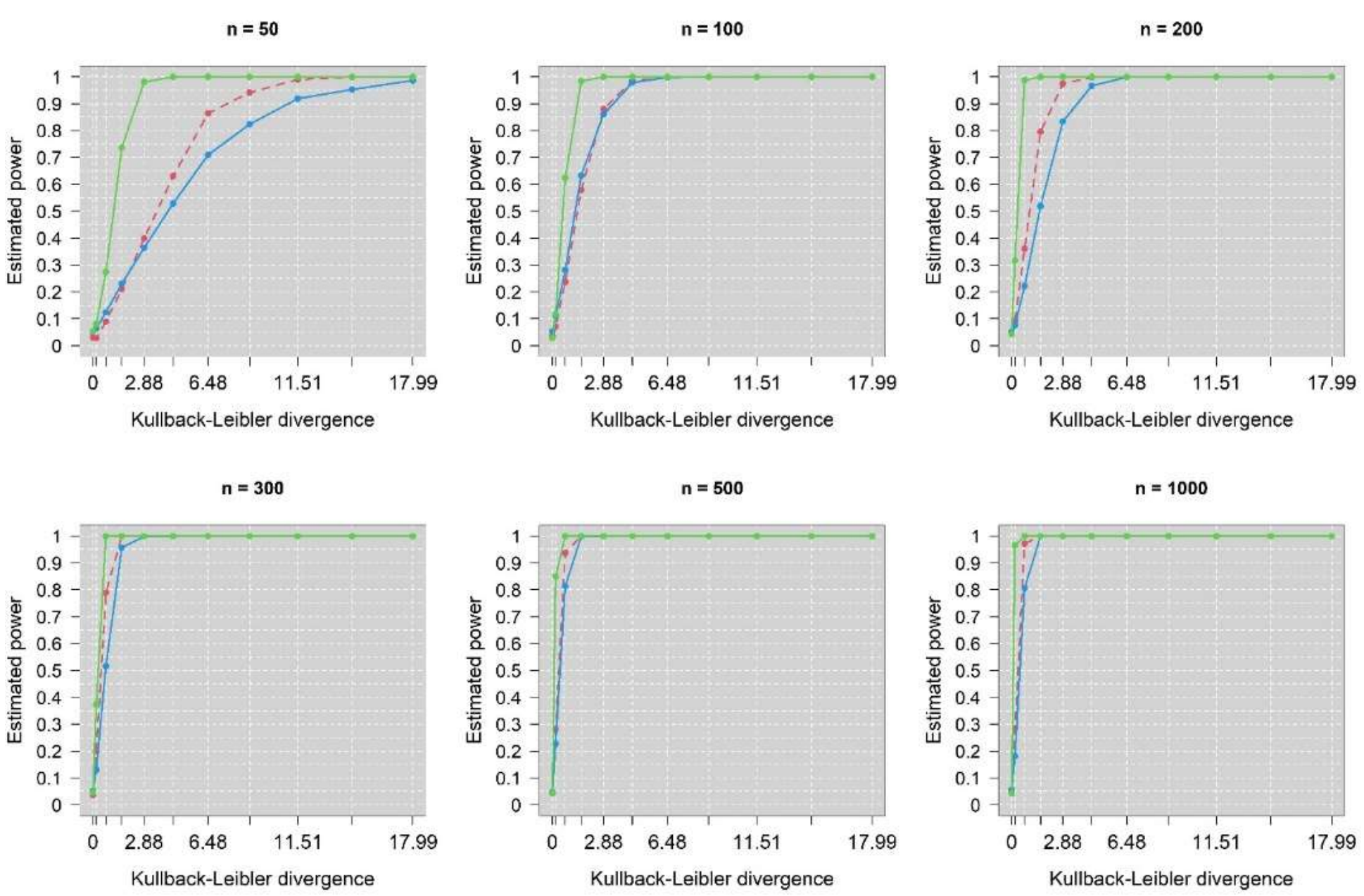

**Figure 2.** Estimated power of the RPBT (red line), and $\alpha$–EBT with $\alpha = 1$ (blue line), and $\alpha$–EBT with $\alpha = 0.1$ (green line) as a function of the KL divergence under Scenario 2.

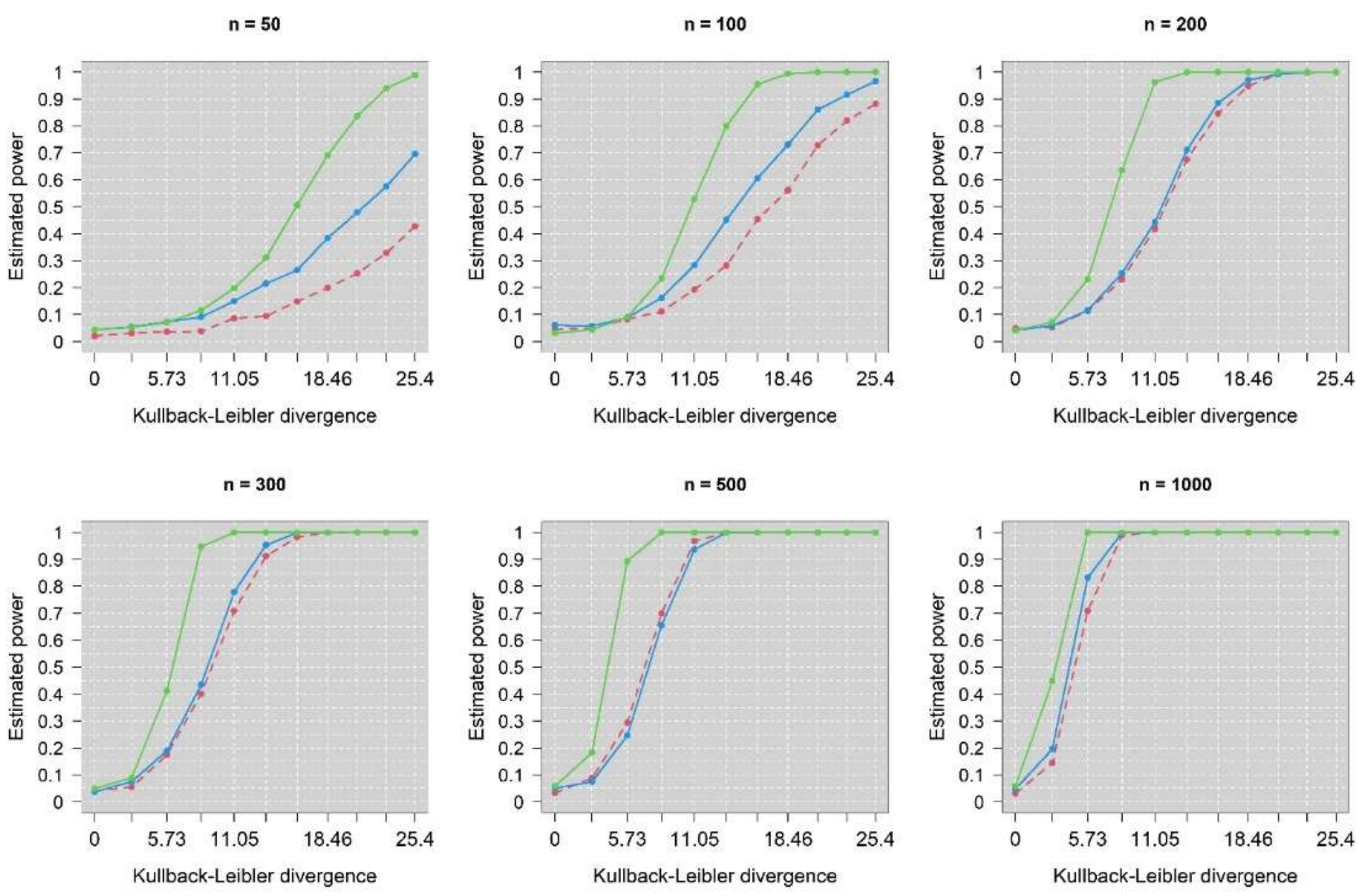

**Figure 3.** Estimated power of the RPBT (red line), $\alpha$–EBT with $\alpha = 1$ (blue line), and $\alpha$–EBT with $\alpha = 0.1$ (green line) as a function of the KL divergence under Scenario 3.

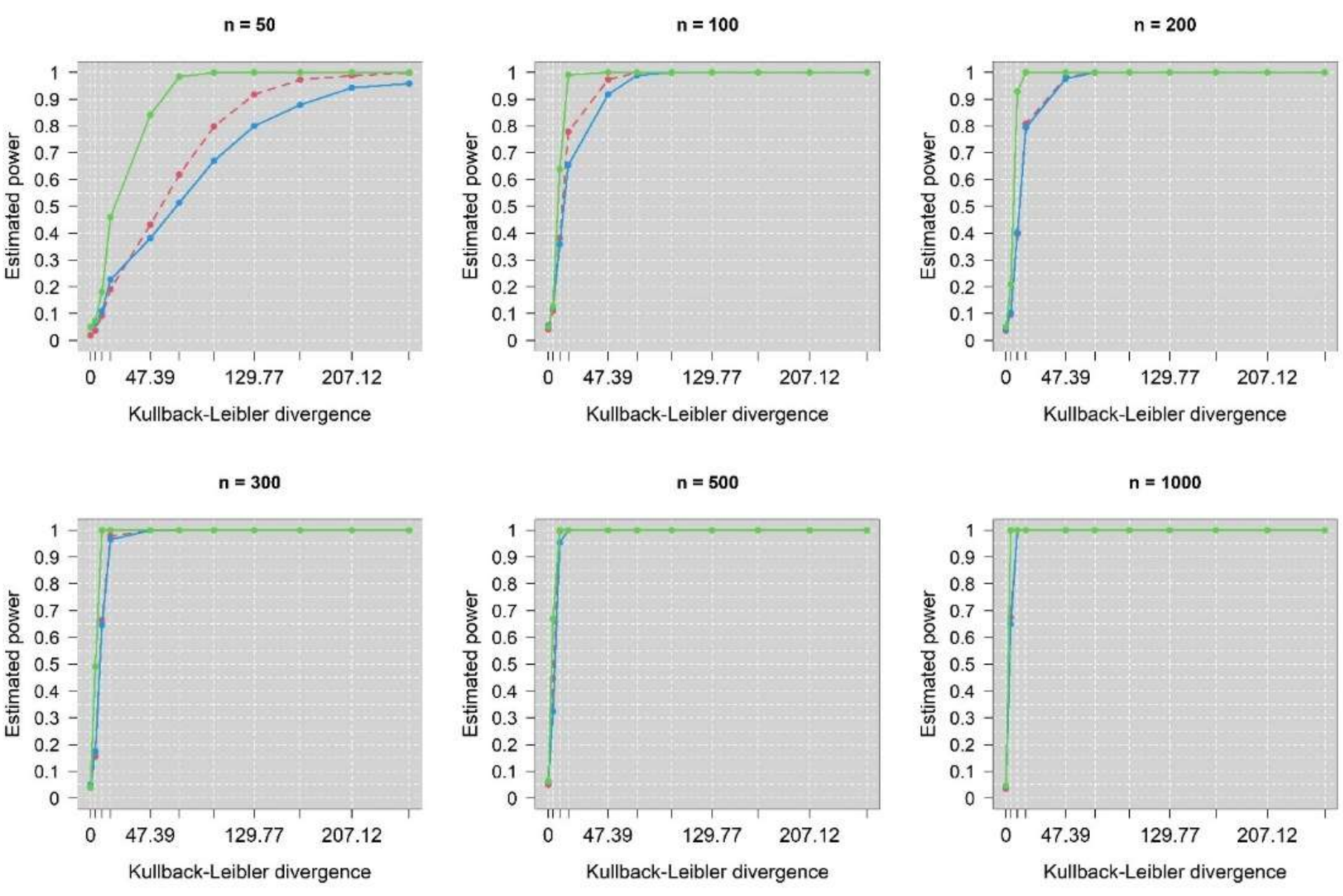

**Figure 4.** Estimated power of the RPBT (red line), $\alpha$–EBT with $\alpha = 1$ (blue line), and $\alpha$–EBT with $\alpha = 0.1$ (green line) as a function of the KL divergence under Scenario 4.

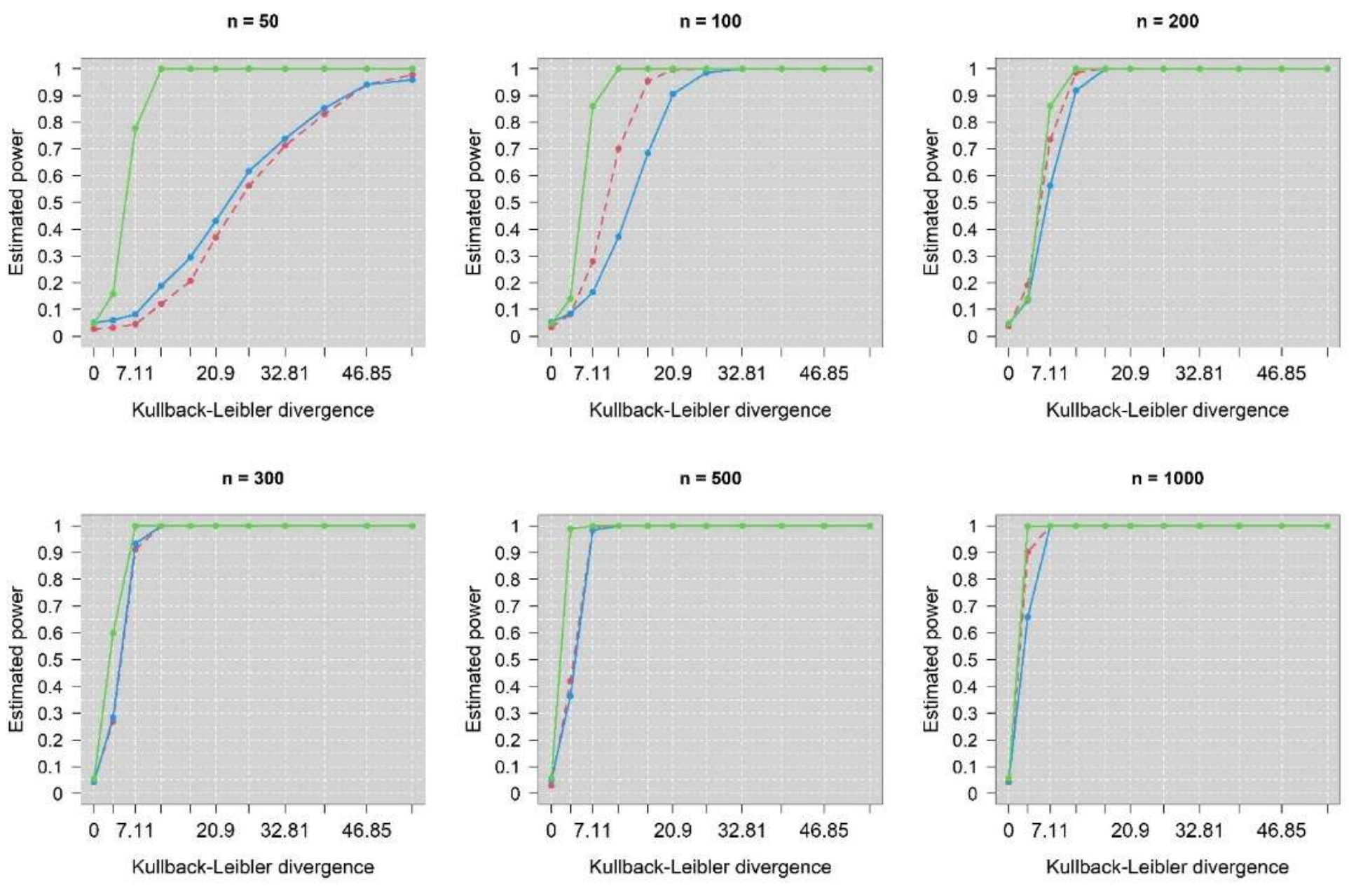

**Figure 5.** Estimated power of the RPBT (red line), $\alpha$–EBT with $\alpha = 1$ (blue line), and $\alpha$–EBT with $\alpha = 0.1$ (green line) as a function of the KL divergence under Scenario 5.

Figure 1 clearly depicts that if the data come from a Dirichlet distribution, the $\alpha$–EBT (regardless of the value of $\alpha$) is significantly more powerful than the RPBT regardless of the sample size. According to Figure 2, the $\alpha$–EBT with $\alpha = 0.1$ is more powerful than the RPBT, which in turn, is more powerful than the $\alpha$–EBT with $\alpha = 1$. Figure 3 shows that the $\alpha$–EBT is more powerful for both values of $\alpha$ when the sample size is equal to 100 or smaller, but for larger sample sizes, the RPBT is more powerful than the $\alpha$–EBT with $\alpha = 1$. The conclusions from Figure 4 are the opposite to those from Figure 3, as the RPBT test is more powerful than the $\alpha$–EBT with $\alpha = 1$ when the sample size is equal to 100 or smaller. However, for larger sample sizes, both tests perform equally well. Finally, in Figure 5, we can observe that depending on the sample size, the RPBT might be better than the $\alpha$–EBT with $\alpha = 1$, but for large sample sizes, they perform almost equally well. However, the common ground for all the case scenarios considered is that the $\alpha$–EBT with $\alpha = 0.1$ always supersedes and reaches the maximum power (equal to 1) always faster than the RPBT or the $\alpha$–EBT with $\alpha = 1$.

One may argue that values of $\alpha$ close to 0 are expected to explain the higher power since the data were generated under the simplicial normal assumption. However, the simplicial normal was used only for Scenarios 2–5, while the first scenario used the Dirichlet distribution following the rationale that $\alpha$–EBT would perform better for values of $\alpha$ close to 1. Evidently, this was not the case. Therefore, this was further explored by adding a sixth case scenario to the existing five ones. Thus, Scenario 6 is a replicate of Scenario 3 but instead of the simplicial normal, the data were generated from the $\alpha$–folded normal (Tsagris and Stewart, 2020) with $\alpha = 1$. The results for the first three sample sizes are presented in Figure 6.

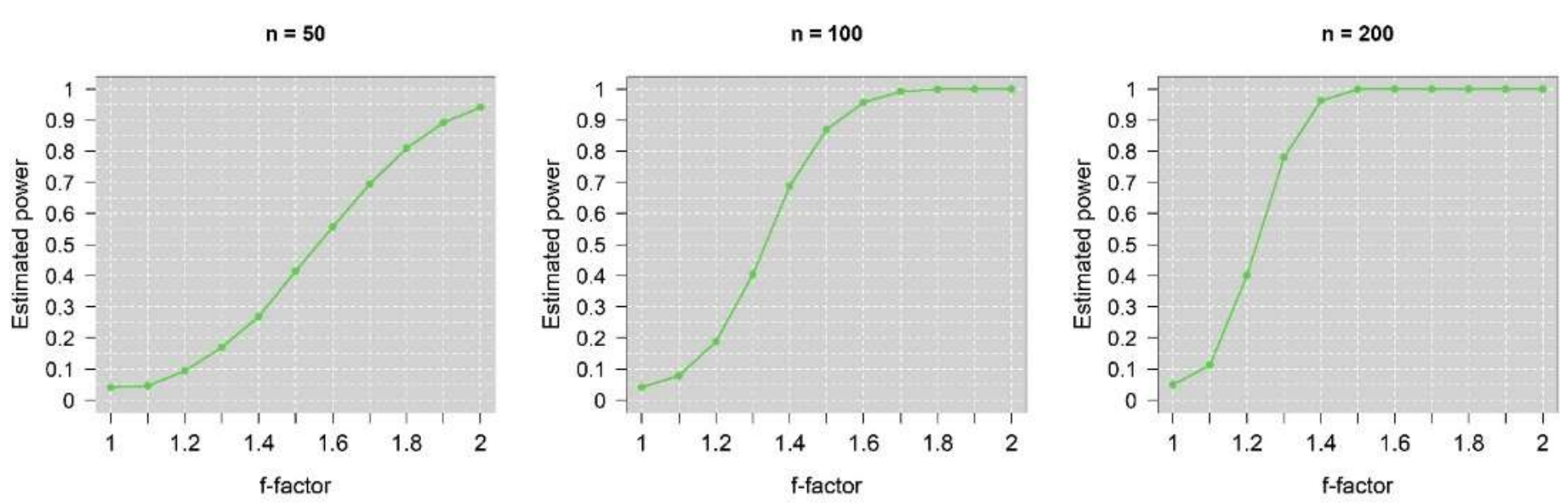

**Figure 6.** Estimated power of the $\alpha$–EBT with $\alpha = 1$ (red line) and $\alpha$–EBT with $\alpha = 0.1$ (green line) as a function of the KL divergence under Scenario 6.

Figure 6 shows that the value of $\alpha$ has no effect on the estimated Type I error probability and the power of the $\alpha$–EBT, as both chosen values of $\alpha$ yield nearly identical results. These results, combined with the previous findings, suggest that the best choice is a small $\alpha$. Since, depending on the situation, a small $\alpha$ could lead to high power levels without compromising the Type I error probability.

A second question naturally raised then is why not use $\alpha = 0$, in which case the $\alpha$–transformation equals the ILR transformation? The answer is that in the case of existing zero values, which is not unlikely to happen with high dimensional data, the ILR would not be applicable. Hence, a safe strategy could be using an even smaller value of $\alpha$, say 0.01. Further, zero replacement strategies could introduce bias in the data. It is possible that future studies could focus on answering this question by performing simulation studies comparing the $\alpha$–EBT with a value of $\alpha = 0.01$ and with a value of $\alpha = 0$ after applying various zero replacement strategies.

## 5. Real Data Analysis

Two real data sets from the field of agricultural economics are used to illustrate the performance of the RPBT and $\alpha$–EBT. The farm accountancy data network (FADN) data supplied by the Greek Ministry of Agriculture regarding crop productivity in the region of Greek NUTS II of Thessaly during the 2017–218 cropping year were used in the analysis. The data refer to a sample of 487 farms. The data set consisted of 20 crops, but after aggregation they were narrowed down to 10 crops. For each of the 487 farms, the cultivated area and the production in each of the 10 crops are known. However, the goal of the paper is to relate the composition of the production to the composition of the cultivated area; and for this reason, the data were scaled to sum to unity[4]. Figure 7 shows the geographical location of the Thessaly region in Greece, and the mean values of each of the 10 components of the FADN data set are provided in Table 2.

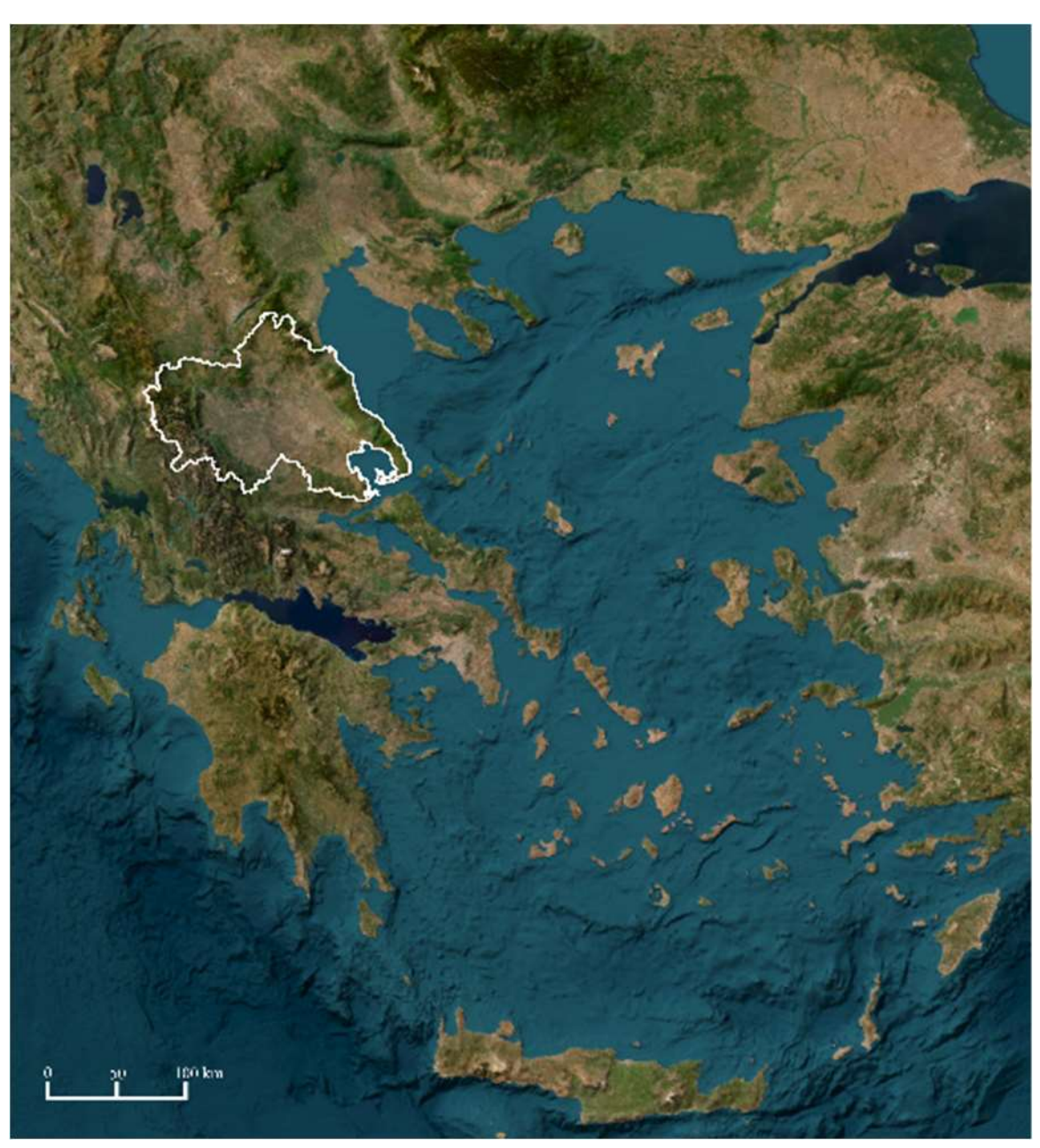

**Figure 7.** The Thessaly region in Greece.

**Table 2.** Average values of the two distributions of the FADN data set.

| Component | Cultivated area | Production |
|---|---|---|
| $X_1$: Other Cereals | 0.1063 | 0.0958 |
| $X_2$: Durum Wheat | 0.1539 | 0.1407 |
| $X_3$: Maize | 0.0600 | 0.1015 |
| $X_4$: Potatoes, Protein Crops and Rice | 0.0393 | 0.0230 |
| $X_5$: Cotton | 0.2315 | 0.2087 |
| $X_6$: Tobacco, Oil Seeds, Industrial Crops and Vegetables | 0.0449 | 0.0708 |

[4] The raw data cannot be distributed due to disclosure restrictions.

| $X_7$: Green Plants, Pasture and Grazing | 0.1777 | 0.1777 |
|---|---|---|
| $X_8$: Fruits, Berries and Nuts | 0.0945 | 0.1098 |
| $X_9$: Olive Trees | 0.0733 | 0.0446 |
| $X_{10}$: Grapes and Wine | 0.0186 | 0.0273 |

When the two tests were applied, the RPBT produced a low p–value ($< 0.001$) indicating that the distributions of the compositions of crop cultivated area and crop production are statistically significantly far from being identical. The $\alpha$–EBT with $\alpha = 0.1$, on the contrary, produced a high p–value ($0.742$) indicating that the two distributions are statistically equivalent. When the tests were applied to the standardized $\alpha$–transformed data, the p–value of the $\alpha$–EBT increased to 1, providing further evidence that the equality of the two distributions could not be rejected. Figure 8 shows the p–values as a function of the $\alpha = (0.1,0.2 \ldots ,0.9,1)$. Evidently, as the $\alpha$ increases the p–values decrease, yet they are always above the 0.05 threshold. Regarding the computational cost, the $\alpha$–EBT required $4.56$ seconds for $\alpha = 0.1$, and $49.36$ seconds for the range of the 10 values of $\alpha$.

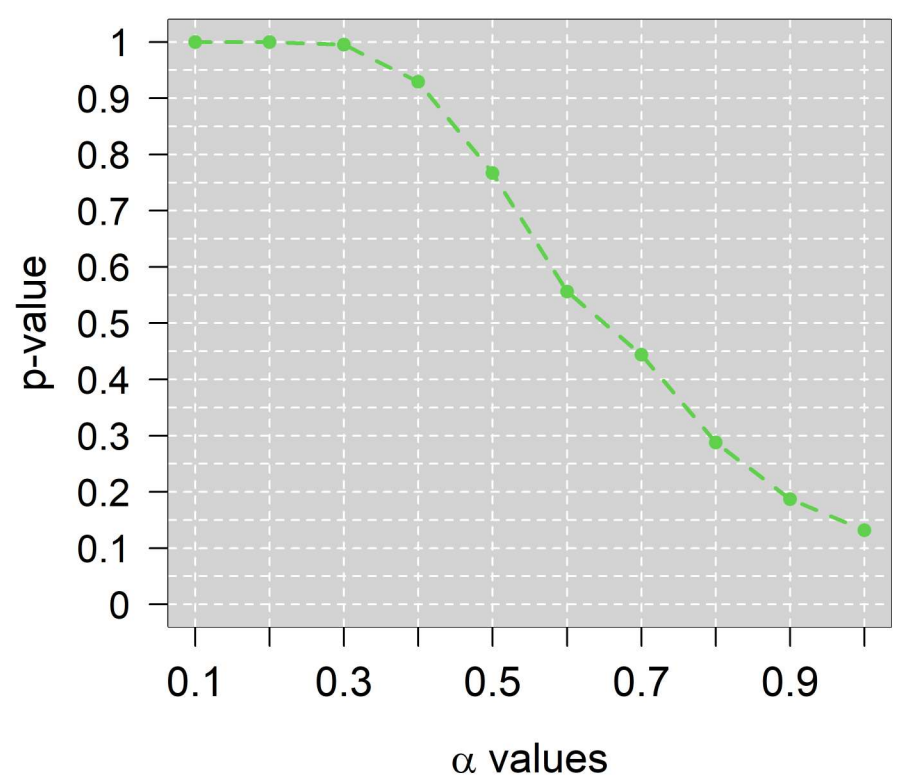

**Figure 8.** The p–values of the $\alpha$–EBT for a range of values of $\alpha$.

## 6. Conclusions

We proposed the novel $\alpha$–EBT test obtained through the adaptation of the energy-based test, which was suggested for equality of distributions of Euclidean data, to the case of compositional data. The adaptation included the $\alpha$–transformation, which was proven to increase the power of the test in some cases. We compared the performance of the $\alpha$–EBT with that of the RPBT, which is the only test proposed for compositional data in the current literature. Our simulation studies showed that unlike the RPBT, the $\alpha$–EBT attains the Type I error probability regardless of the sample size and dimensionality of the simplex.

A potential drawback of the $\alpha$–EBT is the choice of $\alpha$, but our simulation studies have indicated that the safest choice is $\alpha = 1$, and this is the one we propose as well. Another potential disadvantage of $\alpha$–EBT, inherent from the EBT test, is its computational complexity, and hence its running time, especially when working with large datasets. However, our implementation, unlike the EBT (available in the *R* package *energy*) is highly memory efficient. Given its nice statistical properties and its excellent performance, which are evident from the simulation studies, we reach the conclusion that its merits outweigh its computational drawback.

Both the RPBT and the $\alpha$–EBT are available in the *R* package *Compositional*. The function *dptest(x1, x2, B = 100)* performs the RPBT where *x1* and *x2* are the two compositional data sets and *B* denotes the number of random projections. The function *aeqdist.etest(x, sizes, a = 1, R = 999, ms = FALSE)* performs the $\alpha$–EBT, where *x* contains both data sets, *sizes* is a vector that contains the sample sizes of the data sets, *a* denotes the value of $\alpha$, that can be either a single value or a range of values and *R* is the number of permutations to perform in order to compute the p$-$value.

**Statements**

**Author contributions:** M.T. conceptualized the idea and run the simulation studies and real data example, and V.S. wrote the paper. Both authors proof read the paper.

**Data availability:** Not applicable

**Code availability:** Both the RPBT and the $a$–EBT are available in the *R* package *Compositional*.

**Declarations**

**Ethical approval:** This article does not contain any studies with human participants or animals.

**Consent to participate**: The authors consent to participate.

**Consent to publish:** The authors consent to publish.

**Conflict of interest:** The authors declare that there is no conflict of interest related to this study.